\documentclass[a4paper, amsfonts, amssymb, amsmath, reprint, showkeys, nofootinbib, twoside]{revtex4-2}

\usepackage[english]{babel}
\usepackage[utf8]{inputenc}
\usepackage[colorinlistoftodos, color=green!40, prependcaption]{todonotes}
\usepackage{amsthm}
\usepackage{mathtools}
\usepackage{physics}
\usepackage{xcolor}
\usepackage{graphicx}
\usepackage[left=23mm,right=13mm,top=35mm,columnsep=15pt]{geometry} 
\usepackage{adjustbox}
\usepackage{placeins}
\usepackage[T1]{fontenc}
\usepackage{lipsum}
\usepackage{csquotes}
\usepackage{fontawesome}

\usepackage[pdftex, pdftitle={Article}, pdfauthor={Author}]{hyperref} 
\def \githubIcon{{\faGithub}}

\begin{document}

\title{A 3+1 Perturbative Approach to the Cosmic Dynamo Equation}
\author{Juan F. Bravo}
    \email[Correspondence email address: ]{jfbravoc@unal.edu.co}
    \affiliation{Observatorio Astron\'omico Nacional, Facultad de Ciencias, Universidad Nacional de Colombia}
    
    \author{Leonardo Casta\~neda}
    \email[Correspondence email address: ]{lcastanedac@unal.edu.co}
    \affiliation{Observatorio Astron\'omico Nacional, Facultad de Ciencias, Universidad Nacional de Colombia}
    \author{H\'ector J. Hort\'ua}
    \email[Correspondence email address: ]{hhortuao@unbosque.edu.co}
    \affiliation{Grupo Signos, Departamento de Matem\'aticas, Universidad el Bosque, Bogot\'a, Colombia\\Instituto de Neurociencias, Universidad el Bosque, Bogot\'a, Colombia}

\date{\today} 

\begin{abstract}

In this work, we analyze the evolution of PMFs within a perturbed Friedmann-Lema\^{\i}tre-Robertson-Walker (FLRW) spacetime using the formalisms of Numerical Relativity (NR). We apply the 3+1 decomposition to first-order cosmological perturbations to derive the cosmological dynamo equation under the kinematic-dynamo approximation. Our objective is to study the interaction between the seed magnetic field and the growing modes of scalar perturbations, whose associated velocity fields are evolved numerically using the software \texttt{Einstein Toolkit} and \texttt{FLRWSolver}. We find that these velocity fields effectively drive the amplification of the PMF, demonstrating that the extent of this growth is dependent on the electrical conductivity of the cosmic medium. Our findings provide a computational description linking primordial magnetogenesis to the evolution of magnetic seeds, ultimately explaining the ubiquity of large-scale magnetic fields in the universe~\href{https://github.com/FelipeFisMat93/3-1-dynamo-pert-approach}{\githubIcon}.

\end{abstract}


\maketitle

\section{Introduction} \label{sec:Introduction}

In modern cosmology, the precise determination of fundamental parameters is crucial for understanding the universe's structure and evolution \cite{2020DataPlanck}. A significant challenge has emerged in this pursuit, known as the Hubble tension: a persistent and statistically significant discrepancy between the value of the Hubble constant ($H_0$) inferred from early-universe observations, such as the Cosmic Microwave Background (CMB), and that derived from direct, local measurements \cite{DIVALENTINO2025101965, Di_Valentino_2021}. This tension, now exceeding a $5\sigma$ confidence level, suggests that our standard cosmological model, $\Lambda$CDM, may be incomplete, pointing towards the necessity of new physics to reconcile these divergent measurements \cite{universe9020094}.

Within this context, Primordial Magnetic Fields (PMFs) have emerged as a compelling and physically motivated solution \cite{reducing_hubble_tension_sound_horizon, tension2020}. The proposed mechanism posits that the presence of PMFs in the primordial plasma before recombination would induce small-scale inhomogeneities in the baryon density. The resulting magnetic pressure would accelerate the recombination process locally, causing photon decoupling to occur slightly earlier and thus reducing the physical size of the sound horizon at last scattering. A smaller sound horizon, when anchored to the precisely measured angular scale of the CMB, implies a higher inferred value of $H_0$, thereby alleviating the tension with local measurements \cite{schiff2025primordialmagneticfieldsmodified, UCHIDA2025139456, jedamzik2023primordialmagneticfieldshubble}.

A particularly attractive aspect of the PMF scenario is that it is not an \textit{ad hoc} proposal tailored to the Hubble tension. Instead, it is grounded in well-established theoretical frameworks that predate the discovery of this discrepancy. For decades, PMFs have been investigated as a potential explanation for the origin of large-scale magnetic fields observed in galaxies and clusters \cite{galaxies2013, dynamo-galaxies, BEREZHIANI200459}. These fields are thought to originate from high-energy processes in the very early universe, such as inflationary magnetogenesis or cosmological phase transitions, making them a valuable probe of fundamental physics \cite{gurgenidze2025primordialmagneticfieldchiral, Vachaspati_2021, k2018magnetic}.

For PMFs to be considered a viable solution to the Hubble tension and a credible seed for galactic magnetism, it is crucial to understand their evolution throughout cosmic history. A key question is whether seed fields, generated by physical processes in the early universe, can survive and be sufficiently amplified by the dynamics of the expanding cosmos itself \cite{Mtchedlidze_2022, 2831376, Subramanian_2016}. The evolution of these fields is governed by the principles of magnetohydrodynamics (MHD) within a perturbed Friedmann-Lema\^{\i}tre-Robertson-Walker (FLRW) spacetime, where interactions with the cosmic fluid can lead to either dissipation or amplification \cite{durrer2013, Hortúa_2015, repositorioun}.

Addressing this critical question, in this document, we explore the evolution of PMFs using the framework of Numerical Relativity (NR) in a cosmological setting. By applying the 3+1 formalism to a first-order perturbed FLRW metric, we derive the cosmological dynamo equation under the kinematic approximation. This approach allows for a detailed study of how a seed magnetic field interacts with the growing modes of cosmological perturbations. These dynamics are evolved numerically utilizing the \texttt{Einstein Toolkit} \cite{ET-maria_babiuc_hamilton_2019_3522086, einstein-toolkit-zilhao}, a community-driven platform for NR, in conjunction with the \texttt{FLRWSolver} code to set up cosmological initial conditions \cite{PhysRevD.95.064028}.

The principal finding of this analysis is that the primordial magnetic field does indeed undergo amplification, driven by the velocity fields associated with evolving cosmological perturbations. Furthermore, this amplification is dependent on the electrical conductivity of the cosmic medium. This result provides a robust theoretical and computational foundation for the persistence and growth of PMFs, strengthening the hypothesis that a magnetic field of primordial origin could have evolved to the necessary strength to both influence the recombination history and seed the magnetic fields observed in the universe today. This establishes a direct link between the microphysics of the early universe dynamo and the resolution of a major cosmological tension.

Along the text we will use natural units used in \cite{baumgarte_shapiro_book} unless otherwise stated, then $G=c=1$, for the Maxwell equation $\varepsilon_{0}=1$. The tensor indices are given by Greek letters $(\alpha, \beta, \gamma, ...)$ and will take the values from zero (0) to three (3), sometimes Latin indices will be used $(i, j, k, ...)$ and will take the values from one (1) to three (3).
\section{Formalism} \label{sec:develop}
In this section, we will briefly introduce the theoretical background. We will work under the framework of the general theory of relativity (GR). The Einstein field equations without a cosmological constant are given by
\begin{equation}
{}^{4}R_{\alpha\beta}-\frac{1}{2}{}^{4}Rg_{\alpha\beta}=8\pi T_{\alpha\beta},
\label{eq:EinsteinGeneral}
\end{equation}
where $g_{\alpha\beta}$ are the metric tensor components, ${}^{4}R_{\alpha\beta}$ are the four-dimensional Ricci tensor components, and ${}^{4}R$ is the Ricci tensor trace. The Maxwell equations for a curved spacetime are given by
\begin{align}
\nabla_{\left[\alpha\right.}F_{\left.\beta\gamma\right]}&=0,\\
\nabla_{\beta}F^{\alpha\beta}&=4\pi j^{\alpha},
\end{align}
where $\nabla$ is the covariant derivative associated with $g_{\alpha\beta}$. In the following subsections, we will split these equations using two formalisms: the $3+1$ and the $1+3$ formalism. In the third subsection, we will show that it is possible to go from one formalism to the other, which is a manifestation of an equivalence between the $3+1$ and $1+3$ formalisms in the case of the electric and magnetic fields. When both formalisms are introduced in the next pages, it will look like there is a big gap between both of them. This gap is filled in the dynamo section, where this relation is clearer for the goals of this work, especially in the computational implementation.

\subsection{3+1 formulation of GR}
Here, it is possible to take the spacetime and make a foliation composed of a family of spacelike hypersurfaces, allowing us to study the Einstein's equations as a Cauchy problem. In this section, we will follow mostly \cite{eric31, shibata_book}. Let us denote these hypersurfaces by $\Sigma_{t}$, where the parameter of the foliation $t$ corresponds to the coordinate time. From this, we have a normal vector to the hypersurfaces, $\boldsymbol{n}$, which is future-directed and timelike, and a projector to the hypersurfaces, $\boldsymbol{\gamma}$, inducing a spatial metric on each hypersurface. The observers on the hypersurface with a unit timelike vector $\boldsymbol{n}$ are called Eulerian observers. $\boldsymbol{n}$ are tangent to the Eulerian observers' worldlines, and these worldlines are orthogonal to the hypersurfaces. For a timelike $4$-vector $\boldsymbol{t}$ on the spacetime tangent to the time axis, $t^{\alpha}=\left(\partial/\partial t\right)^{\alpha}$ and $t^{\alpha}\nabla_{\alpha}t=1$, we project $\boldsymbol{t}$ along $\boldsymbol{n}$ and $\boldsymbol{\gamma}$, obtaining the lapse function $\alpha$ and the shift vector $\beta^\alpha$. Setting up an induced coordinate system and considering two adjacent hypersurfaces $\Sigma_{t}$ and $\Sigma_{t+\Delta t}$, we are able to write the line element as follows
\begin{multline}
\label{eq:3+1metric}
ds^2 =g_{\mu\nu}dx^\mu dx^\nu\\= -\alpha^2dt^2+\gamma_{ij}\left(dx^i+\beta^idt\right)\left(dx^j+\beta^jdt\right).
\end{multline}

Using $\boldsymbol{\gamma}$ and $\boldsymbol{n}$, we perform the $3+1$ splitting of the energy-momentum tensor
\begin{equation}
T_{\alpha\beta}=En_{\alpha}n_{\beta}+n_{\alpha}p_{\beta}+p_{\alpha}n_{\beta}+S_{\alpha\beta},
\end{equation}
where $E=T_{\alpha\beta}n^{\alpha}n^{\beta}$, $p_{\alpha}=-T_{\mu\nu}\gamma_{\alpha}^{\mu}n^{\nu}$ and $S_{\alpha\beta}=T_{\mu\nu}\gamma_{\alpha}^{\mu}\gamma_{\beta}^{\nu}$. Making a full, perpendicular, and mixed projection onto the hypersurfaces (\ref{eq:EinsteinGeneral}), together with the evolution of $\boldsymbol{\gamma}$, the Einstein field equations in the $3+1$ formalism are given by
\begin{align}
\ & \partial_{t}\gamma_{ij}=\mathcal{L}_{\boldsymbol{\beta}}\gamma_{ij}-2\alpha K_{ij},\label{eq:evol31gamma}\\
\ & \partial_{t}K_{ij}=\mathcal{L}_{\boldsymbol{\beta}}K_{ij}-D_{i}D_{j}\alpha+\alpha\left\{ R_{ij}+KK_{ij}\right. \nonumber \\ \ &\hspace{1.6cm} \left.-2K_{ik}K_{j}^{k}+4\pi\left[\left(S-E\right)\gamma_{ij}-2S_{ij}\right]\right\} ,\label{eq:evol31K}\\
\ & R+K^{2}-K_{ij}K^{ij}=16\pi E,\\
\ & D_{j}K_{i}^{j}-D_{i}K=8\pi p_{i},
\end{align}
where $K_{ij}$ is the extrinsic curvature tensor and $K$ its trace; $D$ is the covariant derivative given by $\boldsymbol{\gamma}$; $R_{ij}$ is the Ricci tensor on the hypersurfaces and $R$ its trace; $E$ is given by the full projection of the energy-momentum tensor $T^{\mu\nu}$ over $\boldsymbol{n}$; $p_i$ is the mixed projection of $T^{\mu\nu}$; $S_{ij}$ is the full spatial projection of $T^{\mu\nu}$, and $S=S_{ij}\gamma^{ij}$.\\
It is also possible to write the Maxwell equations from this formalism. The Faraday tensor $F_{\alpha\beta}$ in this case is given by
\begin{equation}
F_{\alpha\beta}=n_{\alpha}E_{\beta}-E_{\alpha}n_{\beta}+\varepsilon_{\mu\nu\alpha\beta}n^{\mu}B^{\nu},\label{eq:F-eulerian}
\end{equation}
where $\boldsymbol{E}$ and $\boldsymbol{B}$ are the electric and magnetic fields measured by the Eulerian observer, respectively. Therefore, taking the projections along the normal vector $\boldsymbol{n}$ and the induced metric $\boldsymbol{\gamma}$, the Maxwell equations in the $3+1$ formalism are obtained
\begin{align}
&D_iB^i=0,\\
&D_ iE^i=4\pi\rho_e,\\
&\left(\partial_t-\mathcal{L}_{\boldsymbol{\beta}}\right)B^i-\alpha KB^i+\epsilon^{ijk}D_j\left(\alpha E_k\right)=0,\\
 &\left(\partial_t-\mathcal{L}_{\boldsymbol{\beta}}\right)E^i-\alpha KE^i-\epsilon^{ijk}D_j\left(\alpha B_k\right)=-4\pi\alpha J^i,
\end{align}
where we have the divergence-free equation, Gauss, Faraday, and Amp\`ere equations, respectively. In this case, $\epsilon^{\alpha\beta\gamma}=\epsilon^{\alpha\beta\gamma\delta}n_{\delta}$ and
\begin{align}
\rho_e&=-n_\mu j^\mu,\\
J^\alpha&=\gamma^{\alpha\beta}j_{\beta}.
\end{align}

\subsection{1+3 formulation of GR}
In this formulation, we build a set of timelike integral curves with tangent vector $\boldsymbol{u}$. Just like the case of the normal vectors for the Eulerian observers, $\boldsymbol{g}\left(\boldsymbol{u},\boldsymbol{u}\right)=-1$. Observers with velocity vector $\boldsymbol{u}$ are called Lagrangian observers. In this section, we will follow \cite{ellis_maartens_maccallum_2012, 2009ellis-rep-paper}. In this case, similar to the Eulerian observers, $\boldsymbol{u}$ is orthogonal to some hypersurface, but there is not necessarily the same hypersurface for each time. Therefore, the foliation of the spacetime for each observer does not necessarily match at each time. Despite this, it is possible to foliate the spacetime and make a decomposition in terms of $\boldsymbol{u}$ and an orthogonal projector $\boldsymbol{h}$, creating projected symmetric tracefree tensors. This allows us to project the covariant derivative given by the metric tensor $\boldsymbol{g}$ along $\boldsymbol{u}$ and $\boldsymbol{h}$. Let $\boldsymbol{T}$ be a tensor; then
\begin{align}
\bar{\nabla}_{\gamma}T^{\alpha\cdots}_{\beta\cdots}&=h^{\mu}_{\gamma}h^{\alpha}_{\nu}\cdots h^{\sigma}_{\beta}\cdots\nabla_{\mu}T^{\nu\cdots}_{\sigma\cdots},\\
\dot{T}^{\alpha\cdots}_{\beta\cdots}&=u^{\sigma}\nabla_{\sigma}T^{\alpha\cdots},
\end{align}
where we have to keep in mind that $\bar{\nabla}$ is an operator but not necessarily a covariant derivative.\\
This formalism is characterized by a set of kinematic quantities given by the decomposition of the covariant derivative of $\boldsymbol{u}$:
\begin{align}
\nabla_{\beta}u_{\alpha} &=\bar{\nabla}_{\beta}u_{\alpha}-\dot{u}_{\alpha}u_{\beta},\nonumber\\
\ &=\sigma_{\alpha\beta}+\omega_{\alpha\beta}+\frac{1}{3}\Theta h_{\alpha\beta}-\dot{u}_{\alpha}u_{\beta},
\end{align}
where
\begin{align}
\Theta&=\bar{\nabla}_{\alpha}u^{\alpha},\\
\sigma_{\alpha\beta}&=\bar{\nabla}_{\left\langle \beta\right.}u_{\left.\alpha\right\rangle } \nonumber\\
\ &=\left[ h_{\left(\alpha\right.}^{\gamma}h_{\left.\beta\right)}^{\sigma}-\frac{1}{3}h_{\alpha\beta}h^{\gamma\sigma}\right]\bar{\nabla}_{\beta}u_{\alpha}, \\
\omega_{\alpha\beta}&=\bar{\nabla}_{\left[\beta\right.}u_{\left.\alpha\right]}.\\
\end{align}
Here, $\Theta$ represents the volume expansion of a given fluid. The shear tensor $\sigma_{\alpha\beta}$ leaves the volume invariant but determines the distortion arising in the fluid flow. The directions that remain unchanged (principal directions) are eigenvectors of $\sigma_{\alpha\beta}$; other directions are changed. The vorticity tensor $\omega_{\alpha\beta}$ determines a rigid rotation, preserving the relative distances; the magnitude of vorticity is $\sqrt{\omega^{\alpha\beta}\omega_{\alpha\beta}}$. To determine the rotation axis, we define the vorticity vector $\omega^{\delta}=\omega_{\alpha\beta}u_{\gamma}\epsilon^{\alpha\beta\gamma\delta}/2$.
The Maxwell equations can be split into the $1+3$ form relative to $\boldsymbol{u}$. In this case, the Faraday tensor is given by \cite{bona2009elements}
\begin{equation}
F_{\mu\nu}=u_{\mu}e_{\nu}-e_{\mu}u_{\nu}+\epsilon_{\mu\nu\delta\gamma}b^{\delta}u^{\gamma},
\end{equation}
where $\boldsymbol{e}$ and $\boldsymbol{b}$ are the electric and magnetic fields for the Lagrangian observers, respectively. Projected along $\boldsymbol{u}$ and $\boldsymbol{h}$, the Maxwell equations for $1+3$ observers are
\begin{align}
\ & \bar{\nabla}_{\alpha}b^{\alpha}=2\omega^{\alpha}e_{\alpha}, \\
\ & \bar{\nabla}_{\alpha}e^{\alpha}=\rho_{u}-2\omega^{\alpha}b_{\alpha}, \\
\ & h_{\alpha}^{\gamma}\dot{b}^{\alpha}=\left(\sigma_{\beta}^{\gamma}+\omega_{\beta}^{\gamma}-\frac{2}{3}\Theta\delta_{\beta}^{\gamma}\right)b^{\beta}\nonumber\\ \ & \hspace{1.5cm}-\epsilon^{\gamma\mu\nu\beta}u_{\mu}\nabla_{\beta}e_{\nu}-h_{\alpha}^{\gamma}\epsilon^{\mu\nu\alpha\beta}\dot{u}_{\mu}u_{\beta}e_{\nu}, \\
\ & h_{\alpha}^{\gamma}\dot{e}^{\alpha}=\left(\sigma_{\beta}^{\gamma}+\omega_{\beta}^{\gamma}-\frac{2}{3}\Theta\delta_{\beta}^{\gamma}\right)e^{\beta}\nonumber\\ \ &\hspace{1.2cm}+\epsilon^{\gamma\mu\nu\beta}u_{\mu}\nabla_{\beta}b_{\nu}+h_{\alpha}^{\gamma}\epsilon^{\mu\nu\alpha\beta}\dot{u}_{\mu}u_{\beta}b_{\nu}-4\pi J_{u}^{\gamma},
\end{align}
where
\begin{align}
\rho_u&=-u_\mu j^\mu,\\
J_u^\alpha&=h^{\alpha\beta}j_{\beta}.
\end{align}
Including Ohm's law, which is given in the fluid's rest frame \cite{lichnerowicz1967relativistic}
\begin{equation}
J^\alpha_u-\rho_u u^\alpha=\sigma F^{\alpha\beta}u_\beta,
\end{equation}
where $\sigma$ is the electrical conductivity.

\subsection{Equivalence between formalisms}
Let us see how the electric and magnetic fields in both formalisms are related. Here, we take into account that the Faraday tensor in each formalism is written differently. Here, we will use the fact that for the $1+3$ formalism, the electric and magnetic fields are defined as \cite{bona2009elements}
\begin{align}
e^{\mu}&=F^{\nu\mu}u_{\nu},& b^{\mu}&={}^{*}F^{\mu\nu}u_{\nu}
\end{align}
where ${*}F^{\mu\nu}$ is the Hodge dual of $F^{\mu\nu}$. Using (\ref{eq:F-eulerian}) in the above expression, we obtain:
\begin{align}
e^{\mu} & =-WE^{\mu}-\left(E^{\nu}u_{\nu}\right)n^{\mu}+\epsilon^{\delta\gamma\mu\nu}B_{\gamma}n_{\delta}u_{\nu},\label{eq:e_proj_full}\\
b^{\mu} & =WB^{\mu}+\left(B^{\nu}u_{\nu}\right)n^{\mu}+\epsilon^{\delta\gamma\mu\nu}E_{\gamma}n_{\delta}u_{\nu},\label{eq:b_proj_full}
\end{align}
where $W=-\boldsymbol{n}\cdot\boldsymbol{u}$ is the Lorentz factor. Now, we must obtain the $3+1$ decomposition from these fields, projecting (\ref{eq:e_proj_full}) and (\ref{eq:b_proj_full}) along the normal vector and the hypersurfaces. For (\ref{eq:e_proj_full}), we obtain
\begin{align}
e^{\mu}n_{\mu} & =E^{\nu}u_{\nu}, & \gamma_{\mu\nu}e^{\nu} & =-WE_{\mu}+\epsilon_{\mu}^{\delta\gamma\alpha}B_{\gamma}n_{\delta}u_{\alpha},\\
\end{align}
and for (\ref{eq:b_proj_full})
\begin{align}
b^{\mu}n_{\mu} & =-B^{\nu}u_{\nu}, & \gamma_{\mu\nu}b^{\nu} & =WB_{\mu}+\epsilon_{\mu}^{\delta\gamma\alpha}E_{\gamma}n_{\delta}u_{\alpha}.
\end{align}
Under the induced coordinate system over the hypersurfaces,
\begin{align}
e^{\mu}n_{\mu} & =E^{j}u_{j}, & e_{i} & =-WE_{i}+\epsilon_{i}^{jk}B_{j}u_{k},\label{eq:eE}\\
b^{\mu}n_{\mu} & =-B^{j}u_{j}, & b_{i} & =WB_{i}+\epsilon_{i}^{jk}E_{j}u_{k}.\label{eq:bB}
\end{align}
For the case of the current vector,
\begin{align}
\rho & =-W\rho_{u}+J_{u}^{\mu}n_{\mu}, & J_{\mu} & =\rho_{u}\left(\gamma_{\mu\nu}u^{\nu}\right)+\gamma_{\mu\nu}J_{u}^{\nu}.
\end{align}
\section{FLRW Universe} \label{sec:flrw}
Having introduced the $3+1$ and $1+3$ formalisms, we now establish a background solution and apply cosmological perturbation theory. It is possible to perturb quantities in both formalisms; in particular, the Maxwell equations will be perturbed to derive the dynamo equation in the following section.

\subsection{Background universe}
\begin{figure}
\centering
\includegraphics[scale=0.6]{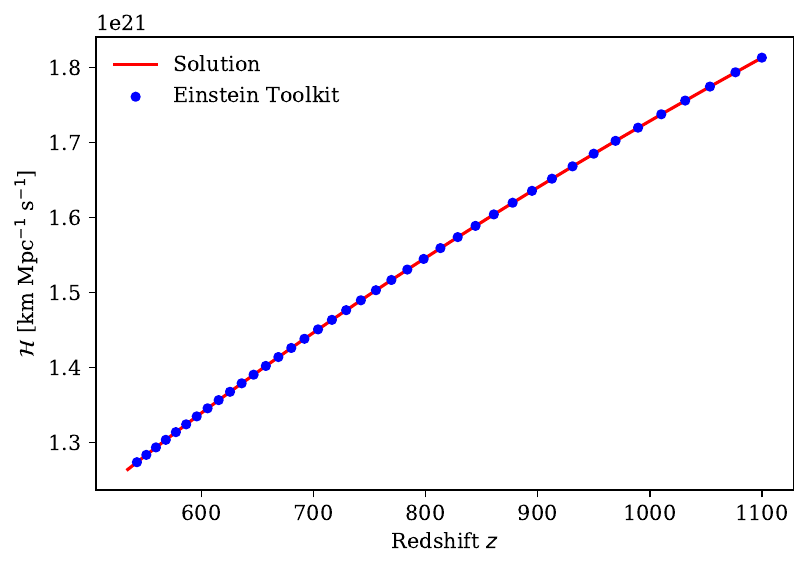}
 \caption{Evolution of the Hubble parameter for the matter-dominated era using the \texttt{Einstein Toolkit}. The red line shows the evolution using expression (\ref{eq:fried-eq}) where $z=-1+\left(z_{\text{CMB}}+1\right)/a$ and $z_{\text{CMB}}=1100$.\label{fig:decayH}}
\end{figure}

To obtain the perturbed equations, we must first define the background spacetime. In this case, we assume a spatially flat Friedmann-Lema\^{\i}tre-Robertson-Walker (FLRW) solution. In this document, we express the line element in terms of conformal time $\tau$, defined such that $d\tau = a^{-1}dt$. This allows us to write the line element as \cite{mukhanov_2005}:
\begin{equation}
ds^{2}=a^{2}\left(\tau\right)\left(-d\tau^{2}+\delta_{ij}dx^{i}dx^{j}\right).\label{eq:flrw-eta}
\end{equation}
Regarding the background evolution, the energy-momentum tensor is modeled as a perfect fluid, whose components are given by:
\begin{equation}
 T_{\beta}^{\alpha}=\left(\rho+p\right)u^{\alpha}u_{\beta}+p\delta_{\beta}^{\alpha},
\end{equation}
where $\rho$ is the energy density and $p$ is the pressure. In the fluid's rest frame, the four-velocity components are:
\begin{equation}
u^{\alpha}=a^{-1}\left(1,0,0,0\right)\text{ and }u_{\beta}=a\left(-1,0,0,0\right).
\end{equation}
From the energy and momentum conservation law, $\nabla_{\mu}T_{\alpha}^{\mu}=0$, setting $\alpha=0$ yields the conservation equation:
\begin{equation}
 \rho'+3\mathcal{H}\left(\rho+p\right)=0,\label{eq:fried-cons}
\end{equation}
where $\mathcal{H}=a'/a$ is the Hubble parameter in conformal time. By setting the $(\alpha=0, \beta=0)$ components of the Einstein equations, the Friedmann equation is obtained\footnote{The background numerical evolution results using the \texttt{Einstein Toolkit} are shown in Figure \ref{fig:decayH}. This background evolution is used in Section \ref{sec:dynamo}.}:
\begin{equation}
 \mathcal{H}^{2}=\frac{8\pi}{3}a^{2}\rho,\label{eq:fried-eq}
\end{equation}
and from the $(i,j)$ components of the energy-momentum tensor, combined with the Friedmann equation, we obtain:
\begin{equation}
 \frac{a''}{a}=\frac{4\pi}{3}a^2\left(\rho-3p\right) + \mathcal{H}^2.\label{eq:a-dos-puntos}
\end{equation}
The solution for equations (\ref{eq:fried-cons}), (\ref{eq:fried-eq}), and (\ref{eq:a-dos-puntos}) is given by \cite{MACPHERSON2019}:
\begin{align}
 a & =a_{0}\xi^{2},\label{eq:a-analytic}\\
\rho & =\rho_{0}\xi^{-6},\label{eq:rho-analytic}\\
\xi & =1+\tau\sqrt{\frac{2}{3}\pi\rho_{0}a_{0}^{2}},\label{eq:zeta-analytic}
\end{align}
where $a_{0}=a\left(\tau_{0}\right)$ and $\rho_{0}=\rho\left(\tau_{0}\right)$ at some initial time $\tau_{0}$. Due to the diffeomorphism invariance of GR, the system of equations (\ref{eq:fried-cons}), (\ref{eq:fried-eq}), and (\ref{eq:a-dos-puntos}) is closed by the equation of state $p\left(\tau\right)=w\left(\tau\right)\rho\left(\tau\right)$.

The total energy density is given by $\rho=\rho_{R}+\rho_{M}$ and the total pressure is $p=p_{R}+p_{M}$, where the indices $R$ and $M$ represent the radiation and matter terms, respectively.

With this in mind, the Friedmann equation can be rewritten as follows:
\begin{equation}
 \mathcal{H}\left(a\right)=\mathcal{H}_{0}\sqrt{\frac{\Omega_{R}}{a^{2}}+\frac{\Omega_{M}}{a}},
\end{equation}
where $\Omega_{R}$ and $\Omega_{M}$ are the respective energy fraction values:
\begin{align}
\Omega_{R} & =\frac{\rho_{R}\left(\tau\right)}{\rho_{c}\left(\tau\right)}, & \Omega_{M} & =\frac{\rho_{M}\left(\tau\right)}{\rho_{c}\left(\tau\right)}.
\end{align}

\subsection{Perturbed universe}
The metric tensor can be split into a background plus a perturbed contribution, $\boldsymbol{g}=\bar{\boldsymbol{g}}+\delta\boldsymbol{g}$. Here, $\bar{\boldsymbol{g}}$ represents the FLRW metric tensor and $\delta\boldsymbol{g}$ denotes the perturbations of the FLRW spacetime. The components of the metric tensor can be expanded as follows \cite{Bruni_1997, Nakamura_2011}:
\begin{align}
g_{00} & =-a^{2}\left(\tau\right)\left(1+2\sum_{n=1}^{\infty}\frac{\psi^{(n)}}{n!}\right),\label{eq:g00-pert}\\
g_{0i} & =a^{2}\left(\tau\right)\sum_{n=1}^{\infty}\frac{\omega_{i}^{(n)}}{n!},\label{eq:g0i-pert}\\
g_{ij} & =a^{2}\left(\tau\right)\left[\left(1-2\sum_{n=1}^{\infty}\frac{\phi^{(n)}}{n!}\right)\delta_{ij}+\sum_{n=1}^{\infty}\frac{\chi_{ij}^{(n)}}{n!}\right],\label{eq:gij-pert}
\end{align}
where $\psi^{(n)}$ and $\phi^{(n)}$ are scalar perturbations, $\omega_{i}^{(n)}$ are vector perturbations, and $\chi_{ij}^{(n)}$ are tensor perturbations, all of order $n$. Writing the metric tensor components $g_{\alpha\beta}$ and its contravariant form $g^{\alpha\beta}$ up to first order in matrix representation, we have:
\begin{align}
 g_{\alpha\beta} & =a^{2}\begin{pmatrix}-\left(1+2\psi\right) & \omega_{i}\\
 \omega_{j} & \left(1-2\phi\right)\delta_{ij}+\chi_{ij}
 \end{pmatrix},\\ g^{\alpha\beta} & =a^{-2}\begin{pmatrix}-\left(1-2\psi\right) & \omega^{i}\\
 \omega^{j} & \left(1+2\phi\right)\delta^{ij}-\chi^{ij}
\end{pmatrix}.\label{eq:pert-metric}
\end{align}
For simplicity, we have removed the index $(1)$; thus, the perturbations are henceforth denoted as $\psi$, $\phi$, $\omega_{i}$, and $\chi_{ij}$. We also consider perturbed matter and electromagnetic quantities for the density $\rho$, pressure $p$, 4-velocity $\boldsymbol{u}$, electromagnetic fields $\boldsymbol{e}$ and $\boldsymbol{b}$, and 4-current $\boldsymbol{j}$:
\begin{align}
 \rho & =\rho_{(0)}+\rho_{(1)},&
 p & =p_{(0)}+p_{(1)},\\
 u^{\alpha} & =\frac{1}{a\left(\tau\right)}\left(\delta_{0}^{\alpha}+v_{(1)}^{\alpha}\right),\label{eq:uvel-pert}&
 e^{i} & =\frac{1}{a^{2}\left(\tau\right)}\left(e_{(0)}^{i}+e_{(1)}^{i}\right),\\
 b^{i} & =\frac{1}{a^{2}\left(\tau\right)}\left(b_{(0)}^{i}+b_{(1)}^{i}\right),&
 j^{i} & =\frac{1}{a\left(\tau\right)}\left(j_{(0)}^{i}+j_{(1)}^{i}\right),
\end{align}
where $v^{\alpha}$ is the peculiar velocity. From the normalization of $\boldsymbol{u}$, we obtain $v_{(1)}^{0}=-\psi$. In the Newtonian gauge, the line element is given by:
\begin{equation}
ds^2=a^{2}\left(\tau\right)\left[-\left(1+2\Psi\right)d\tau^{2}+\left(1-2\Phi\right)\delta_{ij}dx^{i}dx^{j}\right],
\end{equation}
and the field equations are written as follows:
\begin{align}
\nabla^{2}\Phi-3\mathcal{H}\left(\Phi'+\mathcal{H}\Psi\right) & =4\pi\rho\delta a^{2},\label{eq:ham-pert}\\
 \mathcal{H}\partial_{i}\Psi+\partial_{i}\Phi' & =-4\pi\rho a^{2}\delta_{ij}v_{(1)}^{j},\label{eq:mom-pert}\\
 \Phi''+\mathcal{H}\left(\Psi'+2\Phi'\right) & =\frac{1}{2}\nabla^{2}\left(\Phi-\Psi\right),\\
 \left[\partial_{i}\partial_{j}-\frac{1}{3}\delta_{ij}\nabla^{2}\right]\left(\Phi-\Psi\right) & =0,
\end{align}
where $\delta=\rho^{(1)}/\rho$ for the background density $\rho$. Here, (\ref{eq:ham-pert}) and (\ref{eq:mom-pert}) represent the Hamiltonian and momentum constraints, respectively.

\subsection{Perturbed 3+1 formalism}
In this subsection, we apply cosmological perturbation theory to the $3+1$ formalism up to first order. We decompose the lapse function $\alpha$, the shift vector $\boldsymbol{\beta}$, and the induced metric $\boldsymbol{\gamma}$ into their background and first-order perturbation components using (\ref{eq:3+1metric}) and (\ref{eq:pert-metric}), yielding \cite{durrer2013}:
\begin{align}
\alpha^{(0)} & =a\left(\tau\right), & \alpha^{(1)} & =a\left(\tau\right)\psi,\\
 \beta_{i}^{(0)} & =0, & \beta_{i}^{(1)} & =a^{2}\left(\tau\right)\omega_{i},\\
\gamma_{ij}^{(0)} & =a^{2}\left(\tau\right)\delta_{ij}, & \gamma_{ij}^{(1)} & =a^{2}\left(-2\phi\delta_{ij}+\chi_{ij}\right).\label{eq:gamma-pert}\\
\left(n^{\mu}\right)^{(0)} & =a^{-1}\left(1,0,0,0\right), & \left(n^{\mu}\right)^{(1)} & =a^{-1}\left(-\psi,-\omega^{i}\right).
\end{align}
Calculating the extrinsic curvature $K_{ij}$ using (\ref{eq:gamma-pert}) and (\ref{eq:evol31gamma}), we obtain:
\begin{multline}
K_{ij}=-a'\left[\left(1-2\phi+\psi\right)\delta_{ij}+\chi_{ij}\right]\\+a\left[\phi'\delta_{ij}-\frac{1}{2}\chi'_{ij}+\partial_{\left(i\right.}\omega_{\left.j\right)}\right].
\end{multline}
Similarly, we obtain $K^{ij}$ and $K$:
\begin{multline}
K^{ij}=a^{-3}\left[\phi'\delta^{ij}+\mathcal{H}\psi\delta^{ij}-\mathcal{H}\left(1+2\phi\right)\delta_{ij}+\mathcal{H}\chi_{ij}\right.\\ \left.-\frac{1}{2}\chi_{ij}+\frac{1}{2}\left(\delta^{kj}\partial_{k}\omega^{i}+\delta^{ki}\partial_{k}\omega^{j}\right)\right]
\end{multline}
\begin{equation}
K=a^{-1}\left[3\left(-\mathcal{H}+\phi'+\mathcal{H}\psi\right)+\delta^{ij}\partial_{\left(i\right.}\omega_{\left.j\right)}\right].
\end{equation}
The Christoffel symbols for the Levi-Civita connection $D$ are:
\begin{equation}
\Gamma_{ij}^{k}=-2\delta_{\left(i\right.}^{k}\partial_{\left.j\right)}\phi+\partial_{\left(i\right.}\chi_{\left.j\right)}^{k}+\delta_{ij}\delta^{kl}\partial_{l}\phi-\frac{1}{2}\delta^{kl}\partial_{l}\chi_{ij}.
\end{equation}
For the energy-momentum tensor perturbation $T^{\alpha\beta}=T_{(0)}^{\alpha\beta}+T_{(1)}^{\alpha\beta}$, the projections along the hypersurface and the normal vector are:
\begin{align}
 E^{(0)} & =a^{2}T_{(0)}^{00},\\
 p_{i}^{(0)} & =a^{3}T_{(0)}^{0j}\delta_{ij}, \\
 S_{ij}^{(0)} & =a^{4}T_{(0)}^{kl}\delta_{ki}\delta_{lj},
\end{align}
and
\begin{align}
 E^{(1)} & =a^{2}\left(2\psi T_{(0)}^{00}+T_{(1)}^{00}\right),\\
 p_{i}^{(1)} & =a^{3}\left\{ T_{(0)}^{0j}\left[\left(\psi-2\phi\right)\delta_{ij}+\chi_{ij}\right]+T_{(0)}^{00}\omega_{i}+T_{(1)}^{j0}\delta_{ij}\right\}, \\
 S_{ij}^{(1)} & =a^{4}\left\{ T_{(0)}^{kl}\left[\left(\chi_{ki}-2\phi\delta_{ki}\right)\delta_{lj}+\left(\chi_{lj}-2\phi\delta_{lj}\right)\delta_{ki}\right]\right.\nonumber\\&\hspace{2cm} \left.+T_{(1)}^{kl}\delta_{ki}\delta_{lj}\right\} .
\end{align}
In the case of a perfect fluid:
\begin{align}
 E & =\rho_{(0)}+\rho_{(1)},\\
 p_{i} & = a\left(\rho_{(0)}+p_{(0)}\right)\left(v_{i}+\omega_{i}\right),\\
 S_{ij} & =a^{2}p_{(0)}\delta_{ij}+a^{2}\left\{ p_{(0)}\left(\chi_{ij}-2\phi\delta_{ij}\right)+p_{(1)}\delta_{ij}\right\}.
\end{align}
The perturbed $3+1$ Maxwell equations are defined with:
\begin{align}
 B^{i} & =\frac{1}{a^{2}}\left(B_{(0)}^{i}+B_{(1)}^{i}\right), & E^{i}, & =\frac{1}{a^{2}}\left(E_{(0)}^{i}+E_{(1)}^{i}\right).\label{eq:BEup}
\end{align}
The background equations are:
\begin{align}
\ &\partial_{i}B_{(0)}^{i}=0, \\
\ &\partial_{i}E_{(0)}^{i}=4\pi a \rho_{e}^{(0)}, \\
\ & \left(B_{(0)}^{i}\right)^{'}+\mathcal{H}B_{(0)}^{i}+\epsilon^{ij}_{\ k}\partial_{j}E_{(0)}^{k}=0,\\
\ & \left(E_{(0)}^{i}\right)^{'}+\mathcal{H}E_{(0)}^{i}+\epsilon^{ij}_{\ k}\partial_{j}B^{k}_{(0)}=-4\pi a^{2}J_{(0)}^{i}.
\end{align}
The first-order contributions for the magnetic divergence-free, Gauss, Faraday, and Ampère equations are given by:
\begin{align}
 & \partial_{i}B_{(1)}^{i}-3B_{(0)}^{i}\partial_{i}\phi=0, \\
 & \partial_{i}E_{(1)}^{i}-3E_{(0)}^{j}\partial_{i}\phi=0,
\end{align}
\begin{multline}
 \left(B_{(1)}^{i}\right)^{'}+\mathcal{H}B_{(1)}^{i}-B_{(0)}^{i}\left(3\phi'+\delta^{kj}\partial_{\left(j\right.}\omega_{\left.k\right)}\right)\\+B_{(0)}^{k}\partial_{k}\omega^{i}-\omega^{k}\partial_{k}B_{(0)}^{i}+\epsilon^{ij}_{\ k}\left[\partial_{j}E_{(1)}^{k}+E_{(0)}^{k}\partial_{j}\psi\right.\\ \left.+E_{(0)}^{l}\left(\delta_{jl}\delta^{km}\partial_{m}\phi-\frac{1}{2}\delta^{km}\partial_{m}\chi_{jl}\right)\right]=0,\label{eq:faraday3+1-1}
\end{multline}
\begin{multline}
 \left(E_{(1)}^{i}\right)^{'}+\mathcal{H}E_{(1)}^{i}-E_{(0)}^{i}\left(3\phi'+\delta^{kj}\partial_{\left(j\right.}\omega_{\left.k\right)}\right)\\+E_{(0)}^{k}\partial_{k}\omega^{i}-\omega^{k}\partial_{k}E_{(0)}^{i}
-\epsilon^{ij}_{\ k}\left[\partial_{j}B_{(1)}^{k}+\psi\partial_{j}B_{(0)}^{k}-\right.\\
\left.B_{(0)}^{l}\left(\delta_{jl}\partial^{k}\phi-\frac{1}{2}\partial^{k}\chi_{jl}\right)\right]=-4\pi a^{2}\left(J_{(1)}^{i}+\psi J_{(0)}^{i}\right).\label{eq:ampere3+1-1}
\end{multline}

\subsection{Perturbed 1+3 formalism}
In this subsection, we perturb the kinematic quantities and the $1+3$ Maxwell equations. For the velocity vector:
\begin{align}
u^{\mu} & =a^{-1}\left(1-\psi,v_{(1)}^{i}\right),\\
u_{\mu} & =a\left(-1-\psi,\omega_{i}^{(1)}+v_{i}^{(1)}\right).
\end{align}
From the general decomposition of $\nabla_{\beta}u_{\alpha}$, we obtain:
\begin{align}
\omega_{ij} & =a\left(\partial_{\left[i\right.}\omega_{\left.j\right]}^{(1)}+\partial_{\left[i\right.}v_{\left.j\right]}^{(1)}\right),\\
\sigma_{ij} & =a\left(\partial_{\left(i\right.}\omega_{\left.j\right)}^{(1)}+\partial_{\left(i\right.}v_{\left.j\right)}^{(1)}\right)\nonumber\\ \ &-\frac{a^{2}}{3}\left[\left(-\mathcal{H}\psi^{(1)}-\phi^{(1)}\right)\delta_{ij}+\mathcal{H}\left(-2\phi^{(1)}\delta_{ij}+\chi_{ij}^{(1)}\right)\right],\\
\Theta & =3a^{-1}\mathcal{H}+a^{-1}\left(\partial_{j}v^{j}-3\phi'\right).
\end{align}
The background $1+3$ Maxwell equations are:
\begin{align}
\ & \partial_{j}b_{(0)}^{j}=0,\\
\ & \partial_{j}e_{(0)}^{j}=4\pi a \rho_{u}^{(0)},\\
\ & \left(b_{(0)}^{i}\right)'+\mathcal{H}b_{(0)}^{i}+\epsilon^{ij}_{\ k}\partial_{j}e^{k}_{(0)}=0,\label{eq:faraday-b}\\
\ & \left(e_{(0)}^{i}\right)'+\mathcal{H}e_{(0)}^{i}+\epsilon^{ij}_{\ k}\partial_{j}b^{k}_{(0)}=-4\pi a^{2} J_{u(0)}^{i}.\label{eq:ampere-e}
\end{align}
As observed, the $3+1$ and $1+3$ equations are identical in the background, as both formalisms coincide without perturbations. The first-order contributions for the magnetic divergence and Gauss equations are:
\begin{multline}
 \partial_{j}b_{(1)}^{j}+b_{(0)}^{k}\left[-3\partial_{k}\phi-\partial_{k}\psi-2\mathcal{H}\omega_{k}+\frac{1}{2}\partial_{j}\chi_{k}^{j}\right.\\\left.-\frac{1}{2}\delta_{l}^{j}\partial_{l}\chi_{jk}\right]
 +b_{(0)}^{j}\left[a'\left(\omega_{j}+v_{j}\right)+a\left(\omega_{j}+v_{j}\right)'\right]=0,
\end{multline}
\begin{multline}
 \partial_{j}e_{(1)}^{j}+e_{(0)}^{k}\left[-3\partial_{k}\phi-\partial_{k}\psi-2\mathcal{H}\omega_{k}+\frac{1}{2}\partial_{j}\chi_{k}^{j}\right.\\\left.-\frac{1}{2}\delta_{l}^{j}\partial_{l}\chi_{jk}\right]
 +e_{(0)}^{j}\left[a\left(\omega_{j}+v_{j}\right)\right]'=4\pi a \rho_{u}^{(1)},
\end{multline}
The Faraday equation is given by:
\begin{multline}
\left(b_{(1)}^{i}\right)'+\mathcal{H}b_{(1)}^{i}+b_{(0)}^{i}\left[\frac{2}{3}\left(\partial_{k}v^{k}-3\phi'\right)-\psi\mathcal{H}\right]+\\v^{j}\partial_{j}b_{(0)}^{i}
=b_{(0)}^{j}\left\{ a^{-1}\delta^{ik}\left(\partial_{k}\omega_{j}+\partial_{k}v_{j}\right)-\frac{1}{3}\left[\mathcal{H}\left(2\phi\delta_{j}^{i}\right.\right.\right.\\ \left.\left.\left.-\chi_{j}^{i}\right)-\left(\mathcal{H}\psi+\phi'\right)\delta_{j}^{i}\right]-\frac{1}{2}\left(\chi_{j}^{i}\right)'+\phi'\delta_{j}^{i}\right\} \\
-\epsilon^{ij}_{\ k}\left[\partial^{k}e_{j}^{(1)}+e_{(0)}^{l}\left(\partial_{j}\phi\delta^{k}_{l}-\mathcal{H}\omega_{j}\delta_{l}^{k}-\frac{1}{2}\partial_{j}\chi_{l}^{k}\right)\right],\label{eq:fper1}
\end{multline}
and the Ampère equation is:
\begin{multline}
\left(e_{(1)}^{i}\right)'+\mathcal{H}e_{(1)}^{i}+e_{(0)}^{i}\left[\frac{2}{3}\left(\partial_{k}v^{k}-3\phi'\right)-\psi\mathcal{H}\right]\\-\psi \epsilon^{ij}_{\ k}\partial^{k}b_{(0)}^{j}+v^{j}\partial_{j}e_{(0)}^{i}    =e_{(0)}^{j}\left\{ a^{-1}\delta^{ik}\left(\partial_{k}\omega_{j}+\partial_{k}v_{j}\right)\right.\\\left.-\frac{1}{3}\left[\mathcal{H}\left(2\phi\delta_{j}^{i}-\chi_{j}^{i}\right)-\left(\mathcal{H}\psi+\phi'\right)\delta_{j}^{i}\right]-\frac{1}{2}\left(\chi_{j}^{i}\right)'+\phi'\delta_{j}^{i}\right\} \\
-\epsilon^{ij}_{\ k}\left[\partial^{k}b_{j}^{(1)}+b_{(0)}^{l}\left(\partial_{j}\phi\delta^{k}_{l}-\mathcal{H}\omega_{j}\delta^{k}_{l}-\frac{1}{2}\partial_{j}\chi^{k}_{l}\right)\right]\\-4\pi a^{2}\left(\psi J_{u(0)}^{i}+J_{u(1)}^{i}\right).\label{eq:aper1}
\end{multline}
\section{Dynamo equation} \label{sec:dynamo}
In this section, we utilize the perturbed equations derived for the electric and magnetic fields to formulate the dynamo equation. Furthermore, we validate these equations through numerical simulations using the \texttt{Einstein Toolkit} and \texttt{FLRWSolver}.

\subsection{Obtaining the dynamo equation}

We derive the dynamo equation for an FLRW universe from the perspective of a comoving observer ($1+3$ formalism). Consequently, we employ the electromagnetic fields $\boldsymbol{e}$ and $\boldsymbol{b}$. The derivation follows the procedure outlined in \cite{FullDynamo}:
\begin{itemize}
  \item Obtain the curl of $\boldsymbol{b}$ from the Ampère equation.
  \item Apply the curl operator to the curl of $\boldsymbol{b}$.
  \item Use the Faraday, Ampère, and Ohm equations to express terms related to the field $\boldsymbol{e}$ in terms of $\boldsymbol{b}$, then substitute these back into the curl-of-the-curl expression.
\end{itemize}

\begin{figure}[h!]
    \centering
    \includegraphics[scale=0.58]{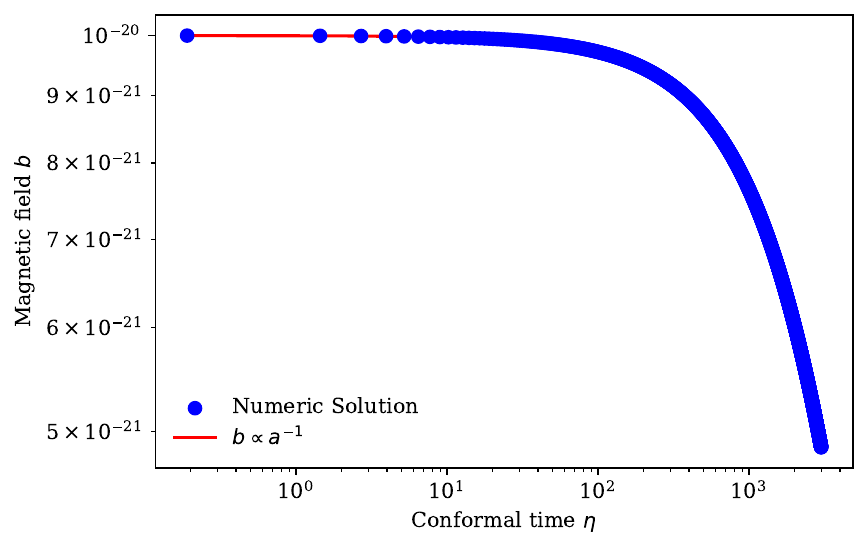}
    \caption{Numerical evolution of Eq. (\ref{eq:dynamo-back}). The decay of the magnetic field is proportional to $a^{-1}$, consistent with a frame choice that allows the field to evolve in this manner \cite{k2018magnetic}.\label{fig:decayb}}
\end{figure}

Starting with the background dynamo equation, we apply the conditions of homogeneity and isotropy to Eq. (\ref{eq:faraday-b}). Letting $b_{(0)}$ represent the magnitude of $b^{i}_{(0)}$, the background dynamo equation is:
\begin{equation}
\label{eq:dynamo-back}
\left(b_{(0)}\right)'+\mathcal{H}b_{(0)}=0.
\end{equation}
As shown in Figure \ref{fig:decayb}, the magnetic field decay is proportional to $a^{-1}$. The background evolution was performed using the \texttt{Einstein Toolkit}, as previously illustrated in Figure \ref{fig:decayH}. While some references suggest a decay of $a^{-2}$ \cite{paper_javier, repositorioun}, it is possible to choose a frame where the field decays as $a^{-1}$, as noted in \cite{k2018magnetic}. 

To obtain the first-order dynamo equation, we combine equations (\ref{eq:fper1}) and (\ref{eq:aper1}). Following several algebraic steps, the first-order dynamo equation is expressed as:
\begin{widetext}
\begin{multline}
    -\partial^{j}\partial_{j}b_{(1)}^{i}=4\mathcal{H}\left[\left(b_{(1)}^{i}\right)'+\mathcal{H}b_{(1)}^{i}\right]-    \left[\left(b_{(1)}^{i}\right)''+\left(\mathcal{H}b_{(1)}^{i}\right)'+b_{(0)}^{j}P_{j}^{i}\right]-\epsilon_{\ k}^{ij}\partial_{j}\left(e_{(0)}^{l}P_{j}^{k}\right)\\
    -\left\{ \partial^{k}e_{k}^{(0)}\left(\omega^{i}+v^{i}-\partial^{i}\psi\right)+e_{k}^{(0)}\partial^{k}\left(\omega^{i}+v^{i}-\partial^{i}\psi\right)-\partial^{j}\left[e^{i}\left(\omega^{i}+v^{i}-\partial^{i}\psi\right)\right]\right\} +\epsilon_{\ k}^{ij}\mathcal{H}\left[e_{l}^{(0)}F_{j}^{l\ k}+\right.\\ 
    \left. e_{k}^{(0)}\left(\omega_{j}+v_{j}-\partial_{j}\psi\right)+\left[F_{j}^{l\ k}+\omega_{j}+v_{j}-\partial_{j}\psi\right]\left(e_{(0)}^{k}\right)'\right]
    +4\pi a^{2}\left\{ \rho^{(0)}v^{i}+\sigma\left[e_{(1)}^{i}+\epsilon^{ijk}\left(\omega_{j}+v_{j}\right)b_{k}^{(0)}\right]\right\} 
\end{multline}
\end{widetext}
where
\begin{align}
    P_{j}^{i} & =\phi'\delta_{j}^{i}-\frac{1}{2}\left(\chi_{j}^{i}\right)'+\partial^{i}\omega_{j}+\partial^{i}v_{j}-\frac{a}{3}\chi_{j}^{i} \nonumber\\
    & \hspace{2.8cm}+\frac{a}{3}\left(\mathcal{H}\psi-\phi+2\mathcal{H}\phi\right)\delta_{j}^{i},
\end{align}
and
\begin{align}
    F_{jk}^{l} & =-2\delta_{\left(j\right.}^{l}\partial_{\left.k\right)}\phi+\delta_{jk}\partial^{l}\phi-\mathcal{H}\omega^{l}\delta_{jk} \nonumber\\
    & \hspace{3.6cm}+\partial_{\left(j\right.}\chi_{\left.k\right)}^{l}-\frac{1}{2}\partial^{l}\chi_{jk}.
\end{align}

Following \cite{paper_javier}, we derive the mean-field dynamo equation. Henceforth, quantities denote Reynolds-averaged values unless otherwise specified \cite{krause1980mean}. At the background level, we assume the magnetic field $b_{(0)}^{i}$ is homogeneous and sufficiently random such that $\left\langle b_{(0)}^{i}\right\rangle = 0$, while $\left\langle b_{i}^{(0)}b_{(0)}^{i}\right\rangle \neq 0$. This implies that terms such as $b_{(0)}^{2}$, $e_{(0)}^{2}$, and $e_{(0)}^{i}$ may be non-zero. Since the universe acts as a conductor at large scales due to coupled charged particles, Ohm's law must be considered. Under these assumptions:
\begin{equation}
    \lim_{\sigma\rightarrow\infty}\frac{J_{u(0)}^{i}}{\sigma}=\lim_{\sigma\rightarrow\infty}e_{(0)}^{i}=0.
\end{equation}
This does not imply a vanishing current; however, a non-zero current would lead to charge separation, breaking background homogeneity. Thus, we set $J_{u(0)}^{i}=0$, and from Gauss's law, $\rho_{u}^{(0)}=0$. Consequently, the only non-zero background term is $b_{(0)}^{2}$, leading to:
\begin{equation}
    b_{(0)}^{i}=e_{(0)}^{2}=e_{(0)}^{i}=J_{u(0)}^{i}=\rho_{u}^{(0)}=0.\label{eq:conditions}
\end{equation}
Under these conditions, the first-order perturbed Ohm's law is:
\begin{equation}
J_{u(1)}^{i}=\sigma\left[e_{(1)}^{i}+\epsilon^{ijk}\left(\omega_{j}+v_{j}\right)b_{k}^{(0)}\right].\label{eq:ohm-pert1}
\end{equation}
The first-order Faraday and Ampère equations in the $1+3$ formalism, (\ref{eq:fper1}) and (\ref{eq:aper1}), simplify to:
\begin{equation}
\left(b_{(1)}^{i}\right)'+\mathcal{H}b_{(1)}^{i}+\epsilon^{ij}_{\ k}\partial_{j}e^{k}_{(1)}=0,\label{eq:far-pert-1}
\end{equation}
\begin{equation}
\left(e_{(1)}^{i}\right)'+\mathcal{H}e_{(1)}^{i}+\epsilon^{ij}_{\ k}\partial_{j}b^{k}_{(1)}=-a^{2}4\pi J_{u(1)}^{i}.\label{eq:amp-pert-1}
\end{equation}
The resulting dynamo equation is:
\begin{widetext}
\begin{multline}
    \label{eq:dinamo1_3}
    \left(b_{(1)}^{i}\right)''=\partial_{j}\partial^{j}b_{(1)}^{i}-\left[2\mathcal{H}+4\pi a^2\sigma\right]\left(b_{(1)}^{i}\right)'-\left[\mathcal{H}'+\mathcal{H}^2+4\pi a^2\sigma\right]b_{(1)}^{i}\\ +4\pi a^2\sigma\partial_j\left[b_{(0)}^j\left(\omega^i+v^i\right)-b_{(0)}^i\left(\omega^i+v^j\right)\right].
\end{multline}
\end{widetext}

To obtain the dynamo equation in the $3+1$ formalism, we use the equivalence relations (\ref{eq:eE}) and (\ref{eq:bB}). Since the electromagnetic fields are restricted to the spatial hypersurface, they lack normal contributions. The perturbed equivalence relations under conditions (\ref{eq:conditions}) are:
\begin{align}
    e_{(1)}^{i} & =E_{(1)}^{i}+a\epsilon_{\ k}^{ij}B_{(0)}^{k}\left(\omega_{j}+v_{j}\right),\\
    b_{(1)}^{i} & =B_{(1)}^{i}+a\epsilon_{\ k}^{ij}E_{(0)}^{k}\left(\omega_{j}+v_{j}\right).
\end{align}
Substituting these into (\ref{eq:dinamo1_3}), the first-order $3+1$ dynamo equation becomes:
\begin{widetext}
\begin{multline}
    \label{eq:dinamo3_1}
    \left(B_{(1)}^{i}\right)''+\epsilon_{\ jk}^{i}\left[aE_{(0)}^{j}\left(v^{k}+\omega^{k}\right)\right]''=    \partial_{j}\partial^{j}\left[B_{(1)}^{i}+\epsilon_{\ k}^{ij}aE_{(0)}^{k}\left(v_{j}+\omega_{j}\right)\right]- \left(2\mathcal{H}+4\pi a^{2}\sigma\right)\left[B_{(1)}^{i}+\epsilon_{\ k}^{ij}aE_{(0)}^{k}\left(v_{j}+\omega_{j}\right)\right]'\\
    -\left(\mathcal{H}'+\mathcal{H}^{2}+4\pi a^{2}\sigma\right)\left[B_{(1)}^{i}+\epsilon_{\ k}^{ij}aE_{(0)}^{k}\left(v_{j}+\omega_{j}\right)\right]+4\pi a^{2}\sigma\partial_{j}\left[b_{(0)}^{j}\left(\omega^{i}+v^{i}\right)-b_{(0)}^{i}\left(\omega^{i}+v^{j}\right)\right].
\end{multline}
\end{widetext}
The primary difference between the two formalisms is the term $aE_{(0)}^{k}\left(v_{j}+\omega_{j}\right)$. We proceed to test Eq. (\ref{eq:dinamo1_3}).

\subsection{Computational implementation}

To test Eq. (\ref{eq:dinamo1_3}), we first employ the $3+1$ formalism within a numerical relativity scheme to obtain the velocity fields that interact with the magnetic fields. Once $v^{i}$ is determined, we evolve the magnetic field via Eq. (\ref{eq:dinamo1_3}). This hybrid approach uses the $3+1$ formalism for the fluid velocity and the $1+3$ formalism for the magnetic field evolution. The growing density modes $\delta$ are computed using the \texttt{Einstein Toolkit} (ET). While ET is standard in astrophysics, the \texttt{FLRWSolver} package enables the setup of cosmological initial conditions to evolve the Einstein field equations \cite{PhysRevD.95.064028, HayleyStruscure}. These modes determine the cosmic fluid's velocity field. In the Newtonian gauge, $\omega^{i}=0$ in Eq. (\ref{eq:dinamo1_3}). Using the kinematic dynamo approximation \cite{rincon_2019, mathDynamos}, we evolve Eq. (\ref{eq:dinamo1_3}) numerically using the method of lines \cite{alcubierrebook}. We found that a first-order spatial difference scheme was necessary for stability, as higher-order schemes caused the simulated system to diverge.

\begin{figure}
    \centering
    \includegraphics[scale=0.58]{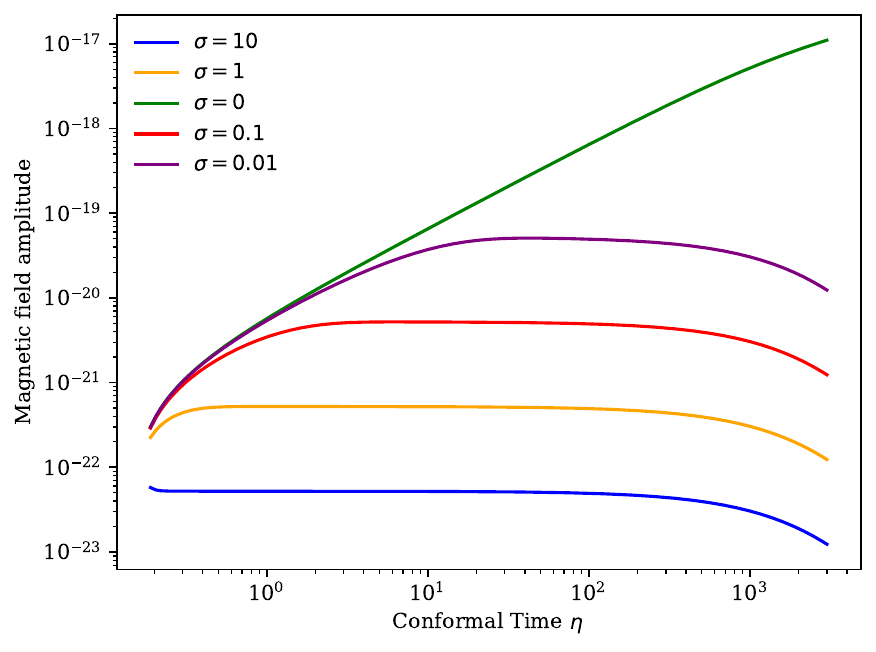}
    \caption{Evolution of the magnetic field amplitude according to Eq. (\ref{eq:dinamo1_3}). The growth is sensitive to the conductivity $\sigma$, starting from an initial magnitude of $B\approx 10^{-22}$. Values of $\sigma$ are given in geometric units.\label{fig:dynamo1_3}}
\end{figure}

The simulations used various conductivity values consistent with galactic environments \cite{JONES2008_DYNAMO}. We also tested the zero-conductivity case ($\sigma=0$), representing cosmic voids. The initial field magnitude was set to $B\approx 10^{-22}$, typical of galaxy clusters \cite{GOVONI_2004}. This order of magnitude separation from the background field was chosen to maintain numerical stability.

\begin{figure} [h]
    \centering
    \includegraphics[scale=0.58]{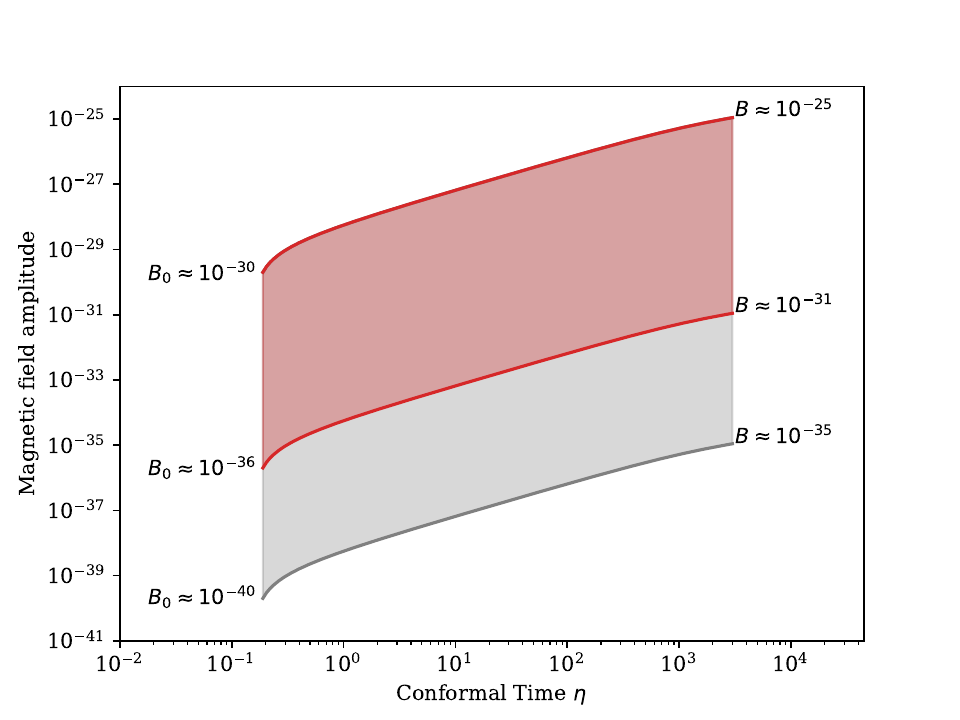}
    \caption{Limits on the magnetic field at recombination based on estimates from \cite{Neronov_2010} and \cite{constrain_pmg_lyman}.\label{fig:voids}}
\end{figure}

Results in Figure \ref{fig:dynamo1_3} show that the magnetic field amplitude grows when $\sigma=0$, as the field interacts with growing modes without dissipation. When $\sigma > 0$, the initial growth is eventually suppressed by dissipation and cosmic expansion, leading to a decay. For $\sigma=0$, the field reached $B\approx 10^{-17}$, whereas higher conductivity led to lower final magnitudes (e.g., $B\approx 10^{-23}$ for $\sigma=10$). 

Finally, the $\sigma=0$ growth model allows us to constrain magnetic fields at recombination. Using current estimates of intergalactic fields ($B \geq 3 \times 10^{-16}$ G \cite{Neronov_2010} and $B \approx 0.2$ nG \cite{constrain_pmg_lyman}), and assuming only linear dynamo effects, our results (Figure \ref{fig:voids}) suggest that initial field values at recombination range from $B \sim 10^{-30}$ to $B \sim 10^{-36}$.
\section{Conclusions} \label{sec:conclusions}

In this paper, we established a methodological framework for studying Primordial Magnetic Fields (PMFs) by reviewing and contrasting the $3+1$ (Eulerian) and $1+3$ (Lagrangian) formalisms. We decomposed the Faraday tensor and Maxwell's equations within each framework, allowing for a direct verification of their mathematical equivalence in describing electromagnetic fields. This unified framework was then applied to a first-order perturbed FLRW spacetime, enabling the derivation of the cosmological dynamo equation in both formalisms and confirming a consistent physical description of PMF evolution.

By employing Numerical Relativity simulations with the \texttt{Einstein Toolkit} and \texttt{FLRWSolver}, we evolved this equation to study the interaction between a seed magnetic field and scalar cosmological perturbations. Our principal finding is that the growing modes of these perturbations, through their associated velocity fields, provide an effective mechanism for the amplification of PMFs. We demonstrated that the magnitude of this amplification depends directly on the electrical conductivity of the cosmic medium. Through this mechanism, we established a method to estimate the field's amplitude following the recombination era, assuming a uniform seed field and the dynamo effect as the primary interaction.

These results provide a robust computational foundation for the PMF hypothesis. By demonstrating a viable pathway for the amplification of a primordial seed field, our work strengthens the theoretical plausibility of PMFs as a potential solution to the Hubble tension. A field amplified by this mechanism could realistically reach the strength necessary to alter recombination history and, consequently, the sound horizon, potentially aligning CMB-inferred $H_0$ values with local measurements.

Future work will build upon this foundation by implementing realistic initial conditions for large-scale, fully non-linear PMF evolution in Numerical Relativity, moving beyond the linear regime explored here. This will enable a direct quantification of how such evolution impacts cosmological observables and allow for a precise comparison with current data related to the Hubble tension. Further avenues of study include investigating the role of decaying modes, extending the analysis to higher-order perturbations, and applying this framework to modified theories of gravity, such as $f(R)$, which are an increasing focus of modern NR studies \cite{PhysRevD.96.024028, figueras2024wellposedinitialvalueformulation, PhysRevD.108.104022, witek_2023_7791842}.
\section*{Acknowledgments} \label{sec:acknowledgements}
We would like to acknowledge the Minister of Science and Technology of Colombia and its project 852 from 2019 called \textit{Physics of the dark universe: Searching for connections
between dark matter, dark energy and particle physics}, which enable us to afford expenses in the University. 

\bibliography{BibliMSc}

@book{eric31,
  title={3+1 Formalism in General Relativity: Bases of Numerical Relativity},
  author={Gourgoulhon, {\'E}.},
  isbn={9783642245244},
  lccn={2011942212},
  series={Lecture Notes in Physics},
  url={https://books.google.com.co/books?id=XwB94Je8nnIC},
  year={2012},
  publisher={Springer Berlin Heidelberg}
}

@book{alcubierrebook,
  title={Introduction to 3+1 Numerical Relativity},
  author={Alcubierre, M.},
  isbn={9780191548291},
  series={International Series of Monographs on Physics},
  url={https://books.google.com.co/books?id=4hDvRvVJeEIC},
  year={2008},
  publisher={OUP Oxford}
}

@book{baumgarte_shapiro_book, place={Cambridge}, title={Numerical Relativity: Solving Einstein's Equations on the Computer}, DOI={10.1017/CBO9781139193344}, publisher={Cambridge University Press}, author={Baumgarte, Thomas W. and Shapiro, Stuart L.}, year={2010}}

@book{shibata_book,
	author = {Shibata, Masaru},
	title = {100 Years of General Relativity Vol. 1: Numerical Relativity},
	publisher = {World Scientific},
	year = {2015},
	doi = {10.1142/9692},
	address = {},
	edition   = {},
	URL = {https://www.worldscientific.com/doi/abs/10.1142/9692},
	eprint = {https://www.worldscientific.com/doi/pdf/10.1142/9692}
}

@article{paper_javier,
  title = {Evolution of magnetic fields through cosmological perturbation theory},
  author = {Hortua, H\'ector J. and Casta\~neda, Leonardo and Tejeiro, J. M.},
  journal = {Phys. Rev. D},
  volume = {87},
  issue = {10},
  pages = {103531},
  numpages = {15},
  year = {2013},
  month = {May},
  publisher = {American Physical Society},
  doi = {10.1103/PhysRevD.87.103531},
  url = {https://link.aps.org/doi/10.1103/PhysRevD.87.103531}
}

@unpublished{repositorioun,
           month = {Octubre},
           title = {Cosmological features of primordial magnetic fields},
          school = {Universidad Nacional de Colombia - Sede Bogot{\'a}},
          author = {H{\'e}ctor J. Hortua},
            year = {2018},
            note = {Doctor en Ciencias - F{\'i}sica. L{\'i}nea de investigaci{\'o}n: F{\'i}sica te{\'o}rica.},
             url = {http://bdigital.unal.edu.co/70714/}
}

@article{PhysRevD.95.064028,
  title = "{Inhomogeneous cosmology with numerical relativity}",
  author = {Macpherson, Hayley J. and Lasky, Paul D. and Price, Daniel J.},
  journal = {Phys. Rev. D},
  volume = {95},
  issue = {6},
  pages = {064028},
  numpages = {13},
  year = {2017},
  month = {Mar},
  publisher = {American Physical Society},
  doi = {10.1103/PhysRevD.95.064028},
  url = {https://link.aps.org/doi/10.1103/PhysRevD.95.064028}
}

@article{MACPHERSON2019,
	author = "Hayley Jessica Macpherson",
	title = "{Inhomogeneous cosmology in an anisotropic Universe}",
	year = "2019",
	month = "9",
	url = "https://bridges.monash.edu/articles/Inhomogeneous_cosmology_in_an_anisotropic_Universe/9790676",
	doi = "10.26180/5d804f40102c7"
}

@book{ellis_maartens_maccallum_2012, place={Cambridge}, title={Relativistic Cosmology}, DOI={10.1017/CBO9781139014403}, publisher={Cambridge University Press}, author={Ellis, George F. R. and Maartens, Roy and MacCallum, Malcolm A. H.}, year={2012}}

@ARTICLE{2009ellis-rep-paper,
       author = {{Ellis}, George F.~R.},
        title = "{Republication of: Relativistic cosmology}",
      journal = {General Relativity and Gravitation},
         year = 2009,
        month = mar,
       volume = {41},
       number = {3},
        pages = {581-660},
          doi = {10.1007/s10714-009-0760-7},
       adsurl = {https://ui.adsabs.harvard.edu/abs/2009GReGr..41..581E}
}

@book{bona2009elements,
  title={Elements of Numerical Relativity and Relativistic Hydrodynamics: From Einstein' s Equations to Astrophysical Simulations},
  author={Bona, C. and Palenzuela-Luque, C. and Bona-Casas, C.},
  isbn={9783642011634},
  lccn={2009926958},
  series={Lecture Notes in Physics},
  url={https://books.google.com.co/books?id=KgPGHaCUaAYC},
  year={2009},
  publisher={Springer Berlin Heidelberg}
}

@book{mukhanov_2005, 
place={Cambridge}, 
title={Physical Foundations of Cosmology}, 
DOI={10.1017/CBO9780511790553}, 
publisher={Cambridge University Press}, 
author={Mukhanov, Viatcheslav}, 
year={2005}}

@book{krause1980mean,
  title={Mean-field Magnetohydrodynamics and Dynamo Theory},
  author={Krause, F. and R{\"a}dler, K.H.},
  isbn={9780080250410},
  lccn={79042947},
  url={https://books.google.com.co/books?id=aoLGAAAAIAAJ},
  year={1980},
  publisher={Pergamon Press}
}

@book{mathDynamos,
  title={Mathematical Aspects of Natural Dynamos},
  author={Dormy, E. and Soward, A.M. (Eds.)},
  isbn={9780429145643},
  year={2007},
  publisher={Chapman and Hall/CRC}
}

@article{rincon_2019, 
title={Dynamo theories}, 
volume={85}, 
DOI={10.1017/S0022377819000539}, 
number={4}, 
journal={Journal of Plasma Physics}, 
publisher={Cambridge University Press}, 
author={Rincon, Fran\c{c}ois}, year={2019}, 
pages={205850401}}

@software{ET-maria_babiuc_hamilton_2019_3522086,
  author       = {Maria Babiuc-Hamilton and et al.},
  title        = "{The Einstein Toolkit}",
  month        = oct,
  year         = 2019,
  note         = {To find out more, visit \url{http://www.einsteintoolkit.org}},
  publisher    = {Zenodo},
  version      = {The "Mayer" release, ET_2019_10},
  doi          = {10.5281/zenodo.3522086},
  url          = {https://doi.org/10.5281/zenodo.3522086}
}

@article{einstein-toolkit-zilhao,
author = {ZILH\~AO, MIGUEL and L\"OFFLER, FRANK},
title = {AN INTRODUCTION TO THE EINSTEIN TOOLKIT},
journal = {International Journal of Modern Physics A},
volume = {28},
number = {22n23},
pages = {1340014},
year = {2013},
doi = {10.1142/S0217751X13400149},

URL = { 
        https://doi.org/10.1142/S0217751X13400149
    
},
eprint = { 
        https://doi.org/10.1142/S0217751X13400149
    
}
}

@article{Bruni_1997,
	doi = {10.1088/0264-9381/14/9/014},
	url = {https://doi.org/10.1088%2F0264-9381%2F14%2F9%2F014},
	year = 1997,
	month = {sep},
	publisher = {{IOP} Publishing},
	volume = {14},
	number = {9},
	pages = {2585--2606},
	author = {Marco Bruni and Sabino Matarrese and Silvia Mollerach and Sebastiano Sonego},
	title = {Perturbations of spacetime: gauge transformations and gauge invariance at second order and beyond},
	journal = {Classical and Quantum Gravity}
}

@article{durrer2013,
author = {Durrer, Ruth and Neronov, Andrii},
year = {2013},
month = {03},
pages = {},
title = {Cosmological Magnetic Fields: Their Generation, Evolution and Observation},
volume = {21},
journal = {The Astronomy and Astrophysics Review},
doi = {10.1007/s00159-013-0062-7}
}

@misc{k2018magnetic,
    title={Magnetic Fields in the Universe},
    author={Kandaswamy Subramanian},
    year={2018},
    eprint={1809.03543},
    archivePrefix={arXiv},
    primaryClass={astro-ph.CO}
}

@article{Subramanian_2016,
doi = {10.1088/0034-4885/79/7/076901},
url = {https://dx.doi.org/10.1088/0034-4885/79/7/076901},
year = {2016},
month = {may},
publisher = {IOP Publishing},
volume = {79},
number = {7},
pages = {076901},
author = {Subramanian, Kandaswamy},
title = {The origin, evolution and signatures of primordial magnetic fields},
journal = {Reports on Progress in Physics}
}

@article{galaxies2013,
   title={Magnetic Fields in Galaxies},
   url={http://dx.doi.org/10.1007/978-94-007-5612-0_13},
   DOI={10.1007/978-94-007-5612-0_13},
   journal={Planets, Stars and Stellar Systems},
   publisher={Springer Netherlands},
   author={Beck, Rainer and Wielebinski, Richard},
   year={2013},
   pages={641–723} }

@article{dynamo-galaxies,
author = {Ruzmaikin, Alexander and Sokolov, Dmitrii and Shukurov, Anvar},
year = {1989},
month = {10},
pages = {1-14},
title = {The dynamo origin of magnetic fields in galaxy clusters},
volume = {241},
journal = {Monthly Notices of the Royal Astronomical Society},
doi = {10.1093/mnras/241.1.1}
}

@article{BEREZHIANI200459,
title = {Generation of large scale magnetic fields at recombination epoch},
journal = {Astroparticle Physics},
volume = {21},
number = {1},
pages = {59-69},
year = {2004},
issn = {0927-6505},
doi = {https://doi.org/10.1016/j.astropartphys.2003.11.002},
url = {https://www.sciencedirect.com/science/article/pii/S0927650503002809},
author = {Z. Berezhiani and A.D. Dolgov}
}

@article{Vachaspati_2021,
  doi = {10.1088/1361-6633/ac03a9},
  url = {https://doi.org/10.1088/1361-6633/ac03a9},
  year = 2021,
  month = {jun},
  publisher = {{IOP} Publishing},
  volume = {84},
  number = {7},
  pages = {074901},
  author = {Tanmay Vachaspati},
  title = {Progress on cosmological magnetic fields},
  journal = {Reports on Progress in Physics}
}

@article{Di_Valentino_2021,
	title={In the realm of the Hubble tension—a review of solutions
	*},
	volume={38},
	ISSN={1361-6382},
	url={http://dx.doi.org/10.1088/1361-6382/ac086d},
	DOI={10.1088/1361-6382/ac086d},
	number={15},
	journal={Classical and Quantum Gravity},
	publisher={IOP Publishing},
	author={Di Valentino, Eleonora and Mena, Olga and Pan, Supriya and Visinelli, Luca and Yang, Weiqiang and Melchiorri, Alessandro and Mota, David F and Riess, Adam G and Silk, Joseph},
	year={2021},
	month=jul, pages={153001} }

@article{tension2020,
  title = {Relieving the Hubble Tension with Primordial Magnetic Fields},
  author = {Jedamzik, Karsten and Pogosian, Levon},
  journal = {Phys. Rev. Lett.},
  volume = {125},
  issue = {18},
  pages = {181302},
  numpages = {6},
  year = {2020},
  month = {Oct},
  publisher = {American Physical Society},
  doi = {10.1103/PhysRevLett.125.181302},
  url = {https://link.aps.org/doi/10.1103/PhysRevLett.125.181302}
}

@incollection{JONES2008_DYNAMO,
title = {Course 2 Dynamo theory},
editor = {Ph. Cardin and L.F. Cugliandolo},
series = {Les Houches},
publisher = {Elsevier},
volume = {88},
pages = {45-135},
year = {2008},
booktitle = {Dynamos},
issn = {0924-8099},
doi = {https://doi.org/10.1016/S0924-8099(08)80006-6},
url = {https://www.sciencedirect.com/science/article/pii/S0924809908800066},
author = {Chris A. Jones}
}

@article{FullDynamo,
    author = {Marklund, M. and Clarkson, C. A.},
    title = "{The general relativistic magnetohydrodynamic dynamo equation}",
    journal = {Monthly Notices of the Royal Astronomical Society},
    volume = {358},
    number = {3},
    pages = {892-900},
    year = {2005},
    month = {04},
    issn = {0035-8711},
    doi = {10.1111/j.1365-2966.2005.08814.x},
    url = {https://doi.org/10.1111/j.1365-2966.2005.08814.x},
    eprint = {https://academic.oup.com/mnras/article-pdf/358/3/892/3415666/358-3-892.pdf},
}

@article{HayleyStruscure,
  title = {Einstein's Universe: Cosmological structure formation in numerical relativity},
  author = {Macpherson, Hayley J. and Price, Daniel J. and Lasky, Paul D.},
  journal = {Phys. Rev. D},
  volume = {99},
  issue = {6},
  pages = {063522},
  numpages = {18},
  year = {2019},
  month = {Mar},
  publisher = {American Physical Society},
  doi = {10.1103/PhysRevD.99.063522},
  url = {https://link.aps.org/doi/10.1103/PhysRevD.99.063522}
}

@article{Nakamura_2011,
	doi = {10.1088/0264-9381/28/12/122001},
  
	url = {https://doi.org/10.1088%2F0264-9381%2F28%2F12%2F122001},
  
	year = 2011,
	month = {may},
  
	publisher = {{IOP} Publishing},
  
	volume = {28},
  
	number = {12},
  
	pages = {122001},
  
	author = {Kouji Nakamura},
  
	title = {General formulation of general-relativistic higher-order gauge-invariant perturbation theory},
  
	journal = {Classical and Quantum Gravity}
}

@book{Nakamura_2020,
	doi = {10.9734/bpi/taps/v3},
  
	url = {https://doi.org/10.9734%2Fbpi%2Ftaps%2Fv3},
  
	year = 2020,
	month = {feb},
  
	publisher = {Book Publisher International (a part of {SCIENCEDOMAIN} International)},
  
	author = {Kouji Nakamura and Anna C. M. Backerra and Heba F. Mansour and Osamu Takagi and Masamichi Sakamoto and Hideo Yoichi and Kimiko Kawano and Mikio Yamamoto and Guido Zbiral},
  
	editor = {Dr. Mohd Rafatullah},
  
	title = {Theory and Applications of Physical Science Vol. 3}
}

@book{lichnerowicz1967relativistic,
  title={Relativistic Hydrodynamics and Magnetohydrodynamics: Lectures on the Existence of Solutions},
  author={Lichnerowicz, A. and Southwest Center for Advanced Studies},
  lccn={67018400},
  series={Mathematical physics monograph series},
  url={https://books.google.com.co/books?id=_gRFAAAAIAAJ},
  year={1967},
  publisher={Benjamin}
}

@article{PhysRevD.96.024028,
	title = {Characteristic formulation for metric $f(R)$ gravity},
	author = {Mongwane, Bishop},
	journal = {Phys. Rev. D},
	volume = {96},
	issue = {2},
	pages = {024028},
	numpages = {14},
	year = {2017},
	month = {Jul},
	publisher = {American Physical Society},
	doi = {10.1103/PhysRevD.96.024028},
	url = {https://link.aps.org/doi/10.1103/PhysRevD.96.024028}
}

@article{GOVONI_2004,
	title={MAGNETIC FIELDS IN CLUSTERS OF GALAXIES},
	volume={13},
	ISSN={1793-6594},
	url={http://dx.doi.org/10.1142/S0218271804005080},
	DOI={10.1142/s0218271804005080},
	number={08},
	journal={International Journal of Modern Physics D},
	publisher={World Scientific Pub Co Pte Lt},
	author={GOVONI, FEDERICA and FERETTI, LUIGINA},
	year={2004},
	month=sep, pages={1549–1594} }

@article{Hortúa_2015,
	doi = {10.1088/0264-9381/32/23/235026},
	url = {https://dx.doi.org/10.1088/0264-9381/32/23/235026},
	year = {2015},
	month = {nov},
	publisher = {IOP Publishing},
	volume = {32},
	number = {23},
	pages = {235026},
	author = {Héctor J. Hortúa and Leonardo Castañeda},
	title = {Contrasting formulations of cosmological perturbations in a magnetic FLRW cosmology},
	journal = {Classical and Quantum Gravity}
}

@article{Neronov_2010,
	title={Evidence for Strong Extragalactic Magnetic Fields from Fermi Observations of TeV Blazars},
	volume={328},
	ISSN={1095-9203},
	url={http://dx.doi.org/10.1126/science.1184192},
	DOI={10.1126/science.1184192},
	number={5974},
	journal={Science},
	publisher={American Association for the Advancement of Science (AAAS)},
	author={Neronov, Andrii and Vovk, Ievgen},
	year={2010},
	month=apr, pages={73–75} }

@article{constrain_pmg_lyman,
	title = {Constraints on Primordial Magnetic Fields from the Lyman-$\ensuremath{\alpha}$ Forest},
	author = {Pavi\ifmmode \check{c}\else \v{c}\fi{}evi\ifmmode \acute{c}\else \'{c}\fi{}, Mak and Ir\ifmmode \check{s}\else \v{s}\fi{}i\ifmmode \check{c}\else \v{c}\fi{}, Vid and Viel, Matteo and Bolton, James S. and Haehnelt, Martin G. and Martin-Alvarez, Sergio and Puchwein, Ewald and Ralegankar, Pranjal},
	journal = {Phys. Rev. Lett.},
	volume = {135},
	issue = {7},
	pages = {071001},
	numpages = {7},
	year = {2025},
	month = {Aug},
	publisher = {American Physical Society},
	doi = {10.1103/77rd-vkpz},
	url = {https://link.aps.org/doi/10.1103/77rd-vkpz}
}

@article{PhysRevD.108.104022,
  title = {Solving the initial conditions problem for modified gravity theories},
  author = {Brady, Sam E. and Arest\'e Sal\'o, Llibert and Clough, Katy and Figueras, Pau and S., Annamalai P.},
  journal = {Phys. Rev. D},
  volume = {108},
  issue = {10},
  pages = {104022},
  numpages = {9},
  year = {2023},
  month = {Nov},
  publisher = {American Physical Society},
  doi = {10.1103/PhysRevD.108.104022},
  url = {https://link.aps.org/doi/10.1103/PhysRevD.108.104022}
}

@software{witek_2023_7791842,
  author       = {Helvi Witek and et al.},
  title        = {Canuda: a public numerical relativity library to
                   probe fundamental physics
                  },
  month        = may,
  year         = 2023,
  publisher    = {Zenodo},
  doi          = {10.5281/zenodo.7791842},
  url          = {https://doi.org/10.5281/zenodo.7791842},
}

@misc{figueras2024wellposedinitialvalueformulation,
      title={Well-posed initial value formulation of general effective field theories of gravity}, 
      author={Pau Figueras and Aaron Held and Áron D. Kovács},
      year={2024},
      eprint={2407.08775},
      archivePrefix={arXiv},
      primaryClass={gr-qc},
      url={https://arxiv.org/abs/2407.08775}, 
}

@misc{jedamzik2023primordialmagneticfieldshubble,
	title={Primordial magnetic fields and the Hubble tension}, 
	author={Karsten Jedamzik and Levon Pogosian},
	year={2023},
	eprint={2307.05475},
	archivePrefix={arXiv},
	primaryClass={astro-ph.CO},
	url={https://arxiv.org/abs/2307.05475}, 
}

@misc{schiff2025primordialmagneticfieldsmodified,
	title={Primordial magnetic fields and modified recombination histories}, 
	author={Jonathan Schiff and Tejaswi Venumadhav},
	year={2025},
	eprint={2506.16517},
	archivePrefix={arXiv},
	primaryClass={astro-ph.CO},
	url={https://arxiv.org/abs/2506.16517}, 
}

@Article{universe9020094,
	AUTHOR = {Hu, Jian-Ping and Wang, Fa-Yin},
	TITLE = {Hubble Tension: The Evidence of New Physics},
	JOURNAL = {Universe},
	VOLUME = {9},
	YEAR = {2023},
	NUMBER = {2},
	ARTICLE-NUMBER = {94},
	URL = {https://www.mdpi.com/2218-1997/9/2/94},
	ISSN = {2218-1997},
	DOI = {10.3390/universe9020094}
}

@article{DIVALENTINO2025101965,
	title = {The CosmoVerse White Paper: Addressing observational tensions in cosmology with systematics and fundamental physics},
	journal = {Physics of the Dark Universe},
	volume = {49},
	pages = {101965},
	year = {2025},
	issn = {2212-6864},
	doi = {https://doi.org/10.1016/j.dark.2025.101965},
	url = {https://www.sciencedirect.com/science/article/pii/S221268642500158X},
	author = {Eleonora Di Valentino and et al.}
}

@Article{reducing_hubble_tension_sound_horizon,
	AUTHOR = {Jedamzik, Karsten and  Pogosian, Levon and Zhao, Gong-Bo},
	TITLE = {Why reducing the cosmic sound horizon alone can not fully resolve the Hubble tension},
	JOURNAL = {Communications Physics},
	VOLUME = {4},
	YEAR = {2021},
	NUMBER = {1},
	ARTICLE-NUMBER = {123},
	URL = {https://doi.org/10.1038/s42005-021-00628-x},
	ISSN = {2399-3650},
	DOI = {10.1038/s42005-021-00628-x}
}

@article{2020DataPlanck,
	title={Planck2018 results: VI. Cosmological parameters},
	volume={641},
	ISSN={1432-0746},
	url={http://dx.doi.org/10.1051/0004-6361/201833910},
	DOI={10.1051/0004-6361/201833910},
	journal={Astronomy and Astrophysics},
	publisher={EDP Sciences},
	author={Aghanim, N. and et al.},
	year={2020},
	month=sep, pages={A6} }

@book{Stewart_1991, place={Cambridge}, series={Cambridge Monographs on Mathematical Physics}, title={Advanced General Relativity}, publisher={Cambridge University Press}, author={Stewart, John}, year={1991}, collection={Cambridge Monographs on Mathematical Physics}}

@unpublished{repositorioun7747,
	month = {Abril},
	title = {Generaci{\'o}n de campos magn{\'e}ticos primordiales / Generation of primordial magnetic fields},
	school = {Universidad Nacional de Colombia},
	author = {H{\'e}ctor Javier  Hortua},
	year = {2011},
	note = {Maestr{\'i}a en Ciencias Astronom{\'i}a},
	url = {http://bdigital.unal.edu.co/7747/}
}

@article{UCHIDA2025139456,
title = {Revisiting constraints on primordial magnetic fields from spectral distortions of cosmic microwave background},
journal = {Physics Letters B},
volume = {865},
pages = {139456},
year = {2025},
issn = {0370-2693},
doi = {https://doi.org/10.1016/j.physletb.2025.139456},
url = {https://www.sciencedirect.com/science/article/pii/S0370269325002175},
author = {Fumio Uchida and Kohei Kamada and Hiroyuki Tashiro}
}

@article{Mtchedlidze_2022,
doi = {10.3847/1538-4357/ac5960},
url = {https://doi.org/10.3847/1538-4357/ac5960},
year = {2022},
month = {apr},
publisher = {The American Astronomical Society},
volume = {929},
number = {2},
pages = {127},
author = {Mtchedlidze, Salome and Domínguez-Fernández, Paola and Du, Xiaolong and Brandenburg, Axel and Kahniashvili, Tina and O’Sullivan, Shane and Schmidt, Wolfram and Brüggen, Marcus},
title = {Evolution of Primordial Magnetic Fields during Large-scale Structure Formation},
journal = {The Astrophysical Journal}
}

@mastersthesis{2831376,
    author = "Mtchedlidze, Salome",
    title = "{Primordial Magnetic Fields: Evolution and Signatures}",
    type = "Other thesis"
}

@misc{gurgenidze2025primordialmagneticfieldchiral,
      title={Primordial magnetic field from chiral plasma instability with sourcing}, 
      author={Murman Gurgenidze and Andrew J. Long and Alberto Roper Pol and Axel Brandenburg and Tina Kahniashvili},
      year={2025},
      eprint={2512.09177},
      archivePrefix={arXiv},
      primaryClass={hep-ph},
      url={https://arxiv.org/abs/2512.09177}, 
}
\bibliographystyle{plain}

\appendix*
\section{Perturbation theory} \label{sec:appendix}
The primary objective of this appendix is to provide a brief introduction to the mathematical foundations of cosmological perturbation theory. The main references for this section are \cite{Nakamura_2020, Nakamura_2011, Bruni_1997, repositorioun7747, paper_javier}.

Mathematically, we consider two distinct spacetimes: the physical spacetime $(\mathcal{M}_{p}, g_{\alpha\beta})$ and the background spacetime $(\mathcal{M}_{0}, \bar{g}_{\alpha\beta})$. A perturbation of any tensor quantity $\boldsymbol{T}$ is defined as the difference between its value in the physical spacetime and its corresponding value in the background spacetime at a given point. To perform this comparison, we require a diffeomorphism $\phi$ between $\mathcal{M}_{0}$ and $\mathcal{M}_{p}$, $\phi: \mathcal{M}_{0} \rightarrow \mathcal{M}_{p}$, which constitutes a gauge choice. This map induces a pullback $\phi^{*}: T_{\phi(p)}^{*}\mathcal{M}_{p} \rightarrow T_{p}^{*}\mathcal{M}_{0}$ for $p \in \mathcal{M}$. Let $\boldsymbol{T}_{0}$ be a tensor defined on $\mathcal{M}_{0}$ and $\boldsymbol{T}$ a tensor defined on $\mathcal{M}_{p}$; the perturbation $\Delta\boldsymbol{T}$ is then defined as:
\begin{equation}
    \Delta\boldsymbol{T}=\left.\phi^{*}\boldsymbol{T}\right|_{\mathcal{M}_{0}}-\boldsymbol{T}_{0},
\end{equation}
where it is understood that this evaluation occurs at each point of $\mathcal{M}_{0}$.

\begin{figure}[h]
    \centering{}\includegraphics{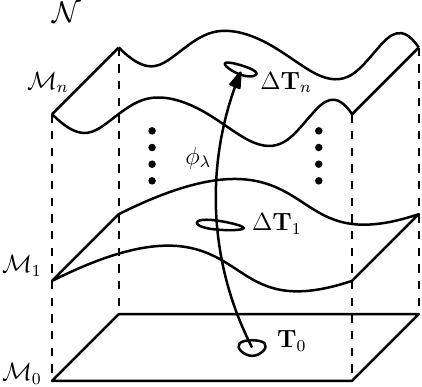}\caption{Scheme of the sub-manifold family $\mathcal{M}_\lambda$ embedded in a five-dimensional manifold $\mathcal{N}$. The comparison between manifolds is established by the flow $\phi_\lambda$.\label{fig:foliation_appendix}}
\end{figure}

To formalize this, let us consider a five-dimensional manifold $\mathcal{N} = \mathcal{M} \times \mathbb{R}$, which contains an embedded family of four-dimensional sub-manifolds $\mathcal{M}_{\lambda}$ indexed by $\lambda \in \mathbb{R}$. A tensor $\boldsymbol{T}_{\lambda}$ residing on $\mathcal{M}_{\lambda}$ can be extended to a tensor $\boldsymbol{T}$ on $\mathcal{N}$ by evaluating it at the point $(p, \lambda)$, where $p \in \mathcal{M}$, such that $\boldsymbol{T}(p, \lambda) = \boldsymbol{T}_{\lambda}(p)$. Each sub-manifold represents a perturbed spacetime, with the background spacetime $\mathcal{M}_{0}$ corresponding to $\lambda = 0$.

To compare a tensor on $\mathcal{M}_{\lambda}$ with its counterpart on $\mathcal{M}_{0}$, we consider a flow $\phi_{\lambda}$, which is the integral curve of a vector field $\boldsymbol{X}$ defined on $\mathcal{N}$. In this five-dimensional space, the vector field components are $\boldsymbol{X} = (X^{0}, X^{1}, X^{2}, X^{3}, X^{4})$, with $X^{4} = 1$ to ensure point mapping between manifolds. By performing a Taylor expansion using $\phi_{\lambda}$, the perturbation is expressed as:
\begin{equation}
    \Delta\boldsymbol{T}_{\lambda}=\left.\phi_{\lambda}^{*}\boldsymbol{T}\right|_{\mathcal{M}_{0}}-\boldsymbol{T}_{0},
\end{equation}
where the expansion is given by:
\begin{equation}
    \left.\phi_{\lambda}^{*}\boldsymbol{T}\right|_{\mathcal{M}_{0}}=\sum_{k=0}^{\infty}\frac{\lambda^{k}}{k!}\delta_{\phi}^{(k)}\boldsymbol{T}=\sum_{k=0}^{\infty}\frac{\lambda^{k}}{k!}\mathcal{L}_{\boldsymbol{X}}^{k}\boldsymbol{T},
\end{equation}
and the $k$-th order perturbation is defined as:
\begin{equation}
    \delta_{\phi}^{(k)}\boldsymbol{T}=\left.\frac{d^{k}}{d\lambda^{k}}\left(\phi_{\lambda}^{*}\boldsymbol{T}\right)\right|_{\lambda=0,\mathcal{M}_{0}}.
\end{equation}
Thus, the total perturbation can be written as the sum:
\begin{equation}
    \Delta\boldsymbol{T}_{\lambda}=\sum_{k=1}^{\infty}\frac{\lambda^{k}}{k!}\delta_{\phi}^{(k)}\boldsymbol{T}.
\end{equation}

Due to the general covariance of General Relativity, we may choose an alternative diffeomorphism $\psi$ between $\mathcal{M}_{0}$ and $\mathcal{M}_{p}$. The choice between different diffeomorphisms is known as a gauge transformation. Let $\psi_{\lambda}$ be another gauge choice, corresponding to the integral curve of a vector field $\boldsymbol{Y}$ (where $\boldsymbol{X} \neq \boldsymbol{Y}$). The expansion under this new gauge is:
\begin{equation}
    \left.\psi_{\lambda}^{*}\boldsymbol{T}\right|_{\mathcal{M}_{0}}=\sum_{k=0}^{\infty}\frac{\lambda^{k}}{k!}\delta_{\psi}^{(k)}\boldsymbol{T}=\sum_{k=0}^{\infty}\frac{\lambda^{k}}{k!}\mathcal{L}_{\boldsymbol{Y}}^{k}\boldsymbol{T},
\end{equation}
where:
\begin{equation}
    \delta_{\psi}^{(k)}\boldsymbol{T}=\left.\frac{d^{k}}{d\lambda^{k}}\left(\psi_{\lambda}^{*}\boldsymbol{T}\right)\right|_{\lambda=0,\mathcal{M}_{0}}.
\end{equation}
A tensor $\boldsymbol{T}$ is considered gauge-invariant if $\phi_{\lambda}^{*}\boldsymbol{T} = \psi_{\lambda}^{*}\boldsymbol{T}$ for any choices $\phi$ and $\psi$. This leads to the Stewart-Walker lemma \cite{Stewart_1991}: for every vector field $\boldsymbol{X}$ and $k \geq 1$,
\begin{equation}
    \mathcal{L}_{\boldsymbol{X}}\delta^{k}\boldsymbol{T}=0,
\end{equation}
if and only if $\boldsymbol{T}$ is gauge-invariant at order $k$. When quantities are not gauge-invariant, we define a gauge transformation between choices as:
\begin{equation}
    \Phi_{\lambda}=\phi_{-\lambda}\circ\psi_{\lambda}.
\end{equation}
This induces a difference between the perturbations $\delta_{\phi}^{(k)}\boldsymbol{T}$ and $\delta_{\psi}^{(k)}\boldsymbol{T}$. For first-order perturbations, this difference is given by:
\begin{equation}
    \delta_{\psi}^{(1)}\boldsymbol{T}-\delta_{\phi}^{(1)}\boldsymbol{T}=\mathcal{L}_{\boldsymbol{\xi}}\boldsymbol{T}_{0},
\end{equation}
where the generator of the gauge transformation is:
\begin{equation}
    \boldsymbol{\xi}=\boldsymbol{Y}-\boldsymbol{X}.
\end{equation}

\end{document}